%% file: main.tex
\documentclass[nocrop]{bioinfo}
\copyrightyear{2025} \pubyear{2025}

\input{header}

\begin{document}

\firstpage{1}

\subtitle{}

\title[MyESL software]{MyESL: Sparse learning in molecular evolution and phylogenetic analysis}
\author[Sanderford et. al. (2025).]{Maxwell Sanderford\,$^{\text{\sfb 1}}$, 
Sudip Sharma\,$^{\text{\sfb 1,2}}$,
Glen Stecher\,$^{\text{\sfb 1}}$,
Jun Liu\,$^{\text{\sfb 3}}$,
Jieping Ye\,$^{\text{\sfb 4}}$,
and Sudhir Kumar\,$^{\text{\sfb 1,2}*}$}
\address{$^{\text{\sf 1}}$Institute for Genomics and Evolutionary Medicine, Temple University, Philadelphia, PA 19122, USA, \\
$^{\text{\sf 2}}$Department of Biology, Temple University, Philadelphia, PA 19122, USA, \\
$^{\text{\sf 3}}$Infinia ML Inc., Durham, North Carolina, USA, and\\
$^{\text{\sf 4}}$Zhejiang Lab, Hangzhou, P.R. China.
}

\corresp{$^\ast$To whom correspondence should be addressed.}

\history{}
\editor{}
\input{abstract}
\maketitle
\input{introduction}
\input{results}
\input{conclusion}

\vspace*{-10pt}
\input{conflict_of_interest}
\vspace*{-10pt}
\input{acknowledgements}
\vspace*{-10pt}
\input{funding}
\vspace*{-10pt}
\input{data_availability}

\bibliographystyle{natbib}
\bibliography{bibliography}

\end{document}

%% file: header.tex
\let\href\undefined 
\usepackage[hidelinks]{hyperref}
 \usepackage{url}
\usepackage{layout}
\usepackage[capitalize]{cleveref}
\usepackage{xspace}
\usepackage{amsmath}
\usepackage{amssymb}
\usepackage{latexsym}  
\usepackage[caption=false]{subfig}  
\usepackage{booktabs}   
\usepackage{colortbl}
\usepackage{dsfont} 
\usepackage{float}
\usepackage{ragged2e}
\usepackage[]{todonotes}  
\usepackage{mathtools}   
\usepackage{graphicx}
\usepackage{epigraph}
\usepackage{siunitx}  
\usepackage{svg}
\usepackage{tikz}
\usetikzlibrary{tikzmark,decorations.pathreplacing,calc}
\usepackage{multirow}   
\usepackage{numprint}   
\npdecimalsign{.}
\usepackage{thmtools}
\usepackage{thm-restate}



%% file: abstract.tex
\abstract{Evolutionary sparse learning (ESL) uses a supervised machine learning approach, Least Absolute Shrinkage and Selection Operator (LASSO), to build models explaining the relationship between a hypothesis and the variation across genomic features (e.g., sites) in sequences alignments. ESL employs sparsity between and within the groups of genomic features (e.g., genomic loci or genes) by using sparse-group LASSO. Although some software packages are available for performing sparse group LASSO, we found them less well-suited for processing and analyzing genome-scale sequence data containing millions of features, such as bases. MyESL software fills the need for open-source software for conducting ESL analyses with facilities to pre-process the input hypotheses and large alignments, make LASSO flexible and computationally efficient, and post-process the output model to produce different metrics useful in functional or evolutionary genomics. MyESL takes binary response or phylogenetic trees as the regression response, processing them into class-balanced hypotheses as required. It also processes continuous and binary features or sequence alignments that are transformed into a binary one-hot encoded feature matrix for analysis. The model outputs are processed into user-friendly text and graphical files. The computational core of MyESL is written in C++, which offers model building with or without group sparsity, while the pre- and post-processing of inputs and model outputs is performed using customized functions written in Python. One of its applications in phylogenomics showcases the utility of MyESL. Our analysis of empirical genome-scale datasets shows that MyESL can build evolutionary models quickly and efficiently on a personal desktop, while other computational packages were unable due to their prohibitive requirements of computational resources and time. MyESL is available for Python environments on Linux and distributed as a standalone application for both Windows and macOS, which can be integrated into third-party software and pipelines.\\
\\
\textbf{Availability:} Download source code, executable, and documentation from \url{https://github.com/kumarlabgit/MyESL}\\
\textbf{Corresponding author:} Sudhir Kumar\href{email:s.kumar@temple.edu}{ (s.kumar@temple.edu).}\\
}

%% file: introduction.tex
\section{Introduction}\label{sec:introduction}
Evolutionary sparse learning (ESL) uses supervised machine learning with a sparsity constraint for comparative sequence analysis in a phylogenetic framework \citep{Kumar2021-rp}. ESL is applied directly to multiple sequence alignments and builds a model for a given phylogenetic hypothesis, such as the grouping of organisms in a clade or the presence or absence of a trait of interest across organisms in a phylogeny. Organisms can be species, individuals, strains, or cells, among other possibilities. ESL model parameters are genomic loci, which can be genes, proteins, exons, introns, intergenic regions, and individual genomic positions (Fig. 1a). 

ESL uses the Least Absolute Shrinkage and Selection Operator, LASSO \citep{Tibshirani1996-zs} and automatically compares alternative models involving different combinations of genomic loci and positions using sparse group lasso with logistic loss \citep{Simon2013-nq, qiao2017systematic}. The selected model reveals key genes and positions containing the most informative shared-derived evolutionary substitutions, along with a measure of the importance of each gene and position referred to as sparsity scores that are larger for more important loci as well as position in those loci \citep{Kumar2021-rp}. The classification model based on these genes could clearly distinguish between members and non-members of a clade in a phylogeny \citep{Kumar2021-rp}. Recently, ESL has been used to detect disruptive sequences and unstable clades in species phylogenies inferred using phylogenomic alignments \citep{sharma2024discovering}. 

The ESL approach was originally implemented using Sparse Learning with
Efficient Projections (SLEP) software \citep{Liu2011-ap}. It implements LASSO regression in MATLAB but does not have built-in functionality to process input sequence alignments and ESL model outputs. MATLAB is neither universally accessible nor free of cost, limiting the use of ESL. For proprietary reasons, the MATLAB version of ESL cannot be distributed with free, user-friendly, and freely available software like MEGA \citep{tamura2021mega11}. In addition, packaging updated MATLAB codes into a standalone software distribution by third parties will require that they have a paid MATLAB compiler. Other open-source computational packages are also available. For example, an R package, \textbf{SGL} \citep{simon2018SGL}, is available for sparse group lasso analysis, incorporating an additional $L1$ sparsity-inducing penalty, but its implementation is computationally intensive and requires substantial memory, making it impractical for large datasets that exceed typical computational resources\citep{liang2022sparsegl}. Other R packages, such as \textbf{gglasso} \citep{yang2020gglasso} and \textbf{biglasso} \citep{zengbiglasso}, use alternative algorithms to improve the speed of group lasso analysis through gradient descent, yet they do not support sparsity at the group level \citep{liang2022sparsegl}. A more recent package, \textbf{sparsegl}, addresses this limitation by providing a fast implementation of sparse group lasso regression with group-level sparsity support \citep{liang2022sparsegl}. However, none of these packages can efficiently handle extremely large datasets with millions of features, such as sequence data, and they lack integrated pre-processing tools for sequence data, making sparse group lasso regression on such datasets infeasible on personal desktops due to excessive computational time or memory requirements (see Section \ref{sub9}). Similarly, the Python package \textbf{agl} \citep{mendez2021agl} faces these same challenges, making sparse group lasso regression on large datasets a significant difficulty.

The above considerations prompted us to develop an open-source software implementing SLEP’s lasso and sparse group lasso functionalities in C\texttt{++} for computational efficiency and facilitating easier updates. We also programmed a new library of key input/output functionalities to process phylogenetic trees and sequence alignments. This library is written in Python and selected for its common use in genomics research. These programming language choices for the new programs, collectively called MyESL, enabled its compilation into platform-specific executables (e.g., MS Windows and macOS) that can be distributed and used without setting up a Python environment. This executable is linked to the MEGA software version 12 \citep{kumar2024mega12}, making its applications accessible via a widely used Graphical User Interface.

%% file: results.tex
\section{Results}\label{sec:results}

The analysis options are provided to the MyESL software on the command line, including a path for the input data file and a text file containing the evolutionary hypothesis. 

\subsection{Reading and processing the evolutionary hypothesis:}\label{subsec1}
MyESL begins by reading the text file containing the evolutionary hypothesis specified in two ways: \verb|--classes <classfile.txt>| or \verb|--tree <treefile.nwk>| directive on the command line. The \verb|classfile.txt| is a tab-separated two-column text file containing the organism name in the first column and a class designation \((+1\) or \(-1)\) in the second column for use as a response in the regression analysis. The +1 class is considered the focal class, a clade in a phylogeny or a trait of interest. 

Alternatively, MyESL can automatically generate classes by processing an input phylogeny provided in a tree file containing a rooted tree in the Newick format. In this tree, the internal node (clade) with a label will be used as the focal clade, i.e., all the names found in the subtree defined by that internal node will be assigned to the +1 class. The remaining taxa in the tree will be assigned to the -1 class. If multiple nodes in the input phylogeny have labels, the \verb|--clade_list <name>| directive specifies the focal clade. MyESL can conduct multiple ESL analyses by automatically generating labels for all the internal nodes in the phylogeny using the \verb|--gen_clade_list| directive, where users are required to define the minimum and maximum count of clade members. 

\begin{figure*}[htb]
    \centering
    \includegraphics[width=0.75\textwidth, height = 1.1\textwidth]{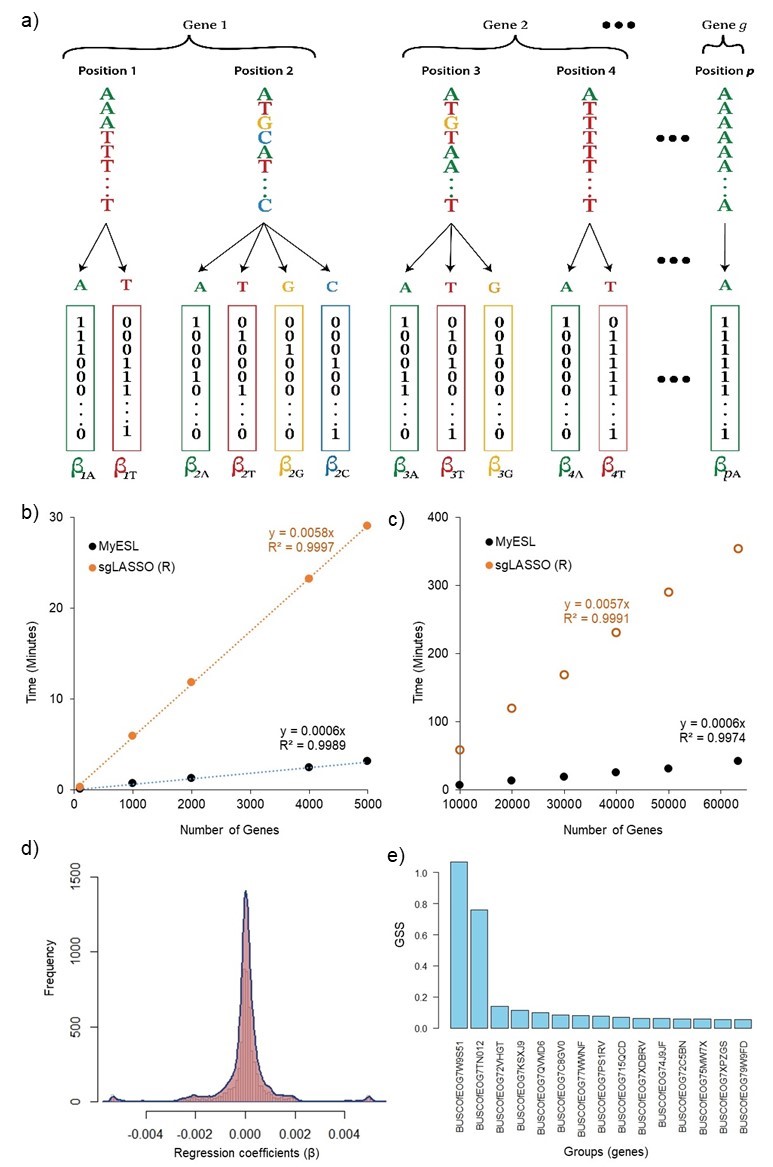}
    \caption{\textbf{One-hot encoding and Evolutionary Sparse Learning.} 
    (a) The sequence alignment input to ESL consists of p positions (columns) belonging to g groups (e.g., genes). The one-hot representation of the alignment is shown below the sequence, where each allele present at a position gets a bit column containing a 1 when the given allele is present in the position and a $0$ otherwise. Every bit column is a feature in the ESL model, which produces weights ($\beta$) for each bit column. $\beta$ captures the correlation between the binary pattern in the bit column and the hypothesis specified by labels ($+1$ or $-1$) assigned to rows in the alignment. (b) Computational time comparison for constructing an ESL model on smaller datasets using MyESL and \textbf{sparagegl}. (c) Computational time comparison for constructing an ESL model on larger datasets using MyESL and \textbf{sparsegl}, where the computational time for \textbf{sparseegl} was projected. (d) The distribution of non-zero regression coefficients estimated in MyESL. (e) Top genes ordered by gene sparsity score (GSS) for features selected in the ESL model.}

    \label{fig:fig1}
    \vspace{-\baselineskip}
    
\end{figure*}

\subsection{Class balancing:}\label{subsec2}

MyESL provides many ways to achieve class balancing, which refers to having an equal number of taxa in the two classes (+1 and -1) using the \verb|--class_bal| command. Two classes can be balanced by up-sampling of the minority class (\verb|<up>| option) or down-sampling of the majority class (\verb|<down>| option). In these two scenarios, the taxa included are selected randomly. The \verb|<weighted>| option will assign weights inversely proportional to the class size to balance the contribution of the two classes. 

MyESL provides a novel option \verb|(<phylo>)| when the evolutionary hypothesis is specified through a Newick tree that contains branch lengths. In this phylogeny-aware class balancing, MyESL first assigns a +1 to all the taxa in the focal clade. Then, it scans the sister of the focal clade and assigns -1 to each taxa therein (i.e., first cousins). If the number of taxa in the -1 class is smaller than the focal clade, then the search for additional taxa continues by moving up to the ancestor of the focal clade and assigning taxa in the sister clade at the next level to be -1 (i.e., second cousins). This process is repeated until no more taxa remain or the number of taxa with a -1 label exceeds those in the focal clade. At this stage, if the number of taxa in two classes is unequal, MyESL prunes the taxon with the shortest terminal branch length in the class with the larger number of taxa, a process repeated until the number of taxa becomes equal between the two classes. In this way, the most diverse taxa are selected for ESL analysis.

\subsection{One-hot encoding of sequence alignment:}\label{subsec3}

Sequence alignments are read from the FastA files. MyESL assumes that each FastA file represents a distinct group of alignment positions, such as a gene or a collection intended to be treated as a group (Fig. 1a). All sequences are then one-hot encoded, a functionality implemented in C++ for maximizing speed. In one-hot encoding, every position in the sequence alignment is converted into as many bit columns as the number of unique characters present at that position. No bit-columns are created for the alignment gap character (\-) and the missing data character (?). 

A \verb|--data_type <nucleotide>| directive informs MyESL to treat A, T, C, G, and U as valid characters without case sensitivity. All other characters will be treated as missing data. Similarly, the \verb|<protein>| option treats all unambiguous IUPAC amino acid letters (case insensitive) as valid characters. The \verb|<molecular>| option provides a way to use both nucleotide and acid letters as valid characters, allowing for the mixing of two data types. By default, however, MyESL will treat all letters (case-sensitive) and digits as distinct characters, which makes MyESL useful for analyzing other types of molecular data. For example, one may include information about the methylation status of positions, the presence/absence of genes or genomic segments, and other molecular characters in the input data files. Even non-molecular characteristics and features of taxa could be specified using letters and digits and then used as input during ESL analysis. When using this flexibility, we suggest users to be careful when preparing their input and interpreting the results. 

MyESL compresses the resulting data matrix by discarding all monomorphic bit-columns across organisms selected for ESL analysis. That is, all the sequences at that position have the same character across taxa, except for the missing data and alignment gaps. Singleton bit-columns are also removed because only one alignment row contains an unambiguous character state distinct from others. Such features will never be informative in the lasso analysis. To further reduce the number of features input to lasso, one may drop all bit-columns in which bit 1 appears fewer than a certain number of times (\verb|--bit_ct <count>|), making MyESL computationally efficient.

\subsection{Model building using LASSO regression}\label{sub4}
MyESL estimates the coefficients of the sparse group LASSO regression model by minimizing the logistic loss \citep{Liu2011-ap}, which is defined as: 
\vspace{-\baselineskip}
\begin{equation}
    L'(\beta) = l(\beta) +  \lambda_{1}|\beta|_{1} + \lambda_{2} \sum\limits_{g=1}^{G} w_{g}\|\beta_{g}\|_{2}.
\end{equation}
\vspace{-\baselineskip}


Here, the first term is the logistic loss function, and the second is the penalty for including individual bit\-columns in the regression model. \(\lambda_{1}\), the regularization parameter, penalizes the inclusion of bit-columns in the ESL model. $\beta$ is the column vector of regression coefficients, with its norm \(\left|\beta\right| = \sum_{i=1}\left|\beta_{i}\right|\) where $i$ ranges from $1$ to the number of bit\-columns in the whole dataset. The third term penalizes the inclusion of groups into the regression model. Here, $\lambda_{2}$ is the group regularization parameter, and \(\|\beta_{g}\| = \sum_{i=1} \left|\beta_{gi}\right|\), where $i$ ranges from $1$ to the number of bit\-columns in group $g$, and $\beta_{gi}$ is the regression coefficient for the $i^{th}$ feature in group $g$. The product of \(\|\beta_{g}\|\) and group weight \((w_{g})\) is summed over all $G$ groups. The group weight is usually the square root of the number of bit columns in group $g$, which MyESL assumes. Alternatively, the user can provide group weights via a text file (\verb|--group_wt <filename.txt>|) containing two tab-separated columns: the first containing group names and the second containing the corresponding group weights. 

In MyESL, $\lambda_{1}$ and $\lambda_{2}$ are the bit-column and group penalty parameters, respectively. These parameters are user-specified by setting  \verb|--lambda1 <float>| and  \verb|--lambda2 <float>| and vary between $0$ and $1$. MyESL also allows performing LASSO regression without group sparsity using the flag \verb|--no_group_penalty| or providing only a single FastA file as data input. 

\subsection{Sparse group lasso implementation}\label{sub5}
The LASSO regression analysis in MyESL is programmed in C\texttt{++} for computational efficiency and portability across platforms. The C\texttt{++} source code is a direct port of the MATLAB code, \verb|sgLogisticR|, for logistic regression with other SLEP functions \cite{Kumar2021-rp}. Regression models optimize the regression weights by employing Moreau-Yosida regularization \citep{Liu2010-fo, Liu2011-ap} and minimize the logistic loss for sparse group logistic lasso regression, respectively. MyESL uses the armadillo library in C\texttt{++} for linear algebra and scientific computing \citep{Sanderson2016-fr}, which can employ multiple computing cores for matrix operations. 

\subsection{ESL Model output, cross-validation, and predictions}\label{sub6}
MyESL uses supervised machine learning to produce an ESL model for the given hypothesis. In this process, ESL estimates regression coefficients \(\beta's\) for each bit-position in a one-hot encoded matrix, where most of these coefficients will be zero \((\beta = 0)\) due to sparsity constraint. MyESL has a Python function (\verb|MyESL_model_apply.py|) to utilize a pre-trained ESL model, specifically, the feature (bit columns) weights or regression coefficients produced. This function can classify a new set of organisms whose sequences are aligned with the data used to build the model. The classification can be performed using logistic loss when the ESL model is built. Each test organism is assigned a prediction score and a probability (0 to 1) output in a tab-separated text. A prediction score greater than zero or a probability greater than 0.5 indicates classification in the class labeled $+1$. 

Users can also build a pre-trained ESL model for classification by performing cross-validation. MyESL optimizes regression weights for multiple training sets and validates the classification accuracy using the held-out sets. The cross-validation in MyESL is performed using the directive \verb|--kfold <int>|. For example, $80\%$ of the taxa are used in model training, while $20\%$ of taxa are withheld for validation if k-fold is set at 5. MyESL produces feature weights from the training samples and classification accuracy for each holdout sample, and users can choose the model with the highest accuracy or use other criteria (e.g., root mean square error). Cross-validation in MyESL can select the best pair of sparsity parameters \citep{Chetverikov2016-pm} or assess model accuracy without using test data \citep{Xu2018-sr}. One should avoid cross-validation if the focal clade has only a few members.

\subsection{Building multiple ESL models}\label{sub7}
Building multiple models with the same feature/response data is a common practice in machine learning to select an optimal pair of sparsity parameters or achieve model averaging. MyESL allows building multiple ESL models by performing a grid search over the regularization parameter space. The grid search option lets users specify the bit (\verb|--lamba1_grid <float, float, float>|) and group (\verb|--lamba1_grid <float, float, float>|) sparsity parameters by defining the minimum [0-1], maximum [0-1], and step size [0-1] of the parameter space. MyESL can prematurely terminate the grid search process to avoid building overly sparse models that may result in high sparsity parameter values. The \verb|--min_group_ct <int>| option sets the minimum number of groups threshold to set the limit of grid search.  

\subsection{ESL output of evolutionary parameters}\label{sub8}
MyESL processes the regression coefficient and produces a series of result files containing different sparsity scores. These are tab-separated text files generated using the following directive \verb|--stats_out <PGHS>|. Different letters in the input string for the directive will produce the corresponding results as follows: P: Position Sparsity Scores; G: Group Sparsity Scores; H: Hypothesis Sparsity Scores; S: A file containing both Species Prediction Score (SPS) and Species Prediction Probability (SPP) (see details in \cite{Kumar2021-rp}).

\subsection{Computational efficiency of MyESL}\label{sub9}
We evaluated the computational efficiency of MyESL against the widely-used R package for sparse group lasso analysis, \textbf{sparsegl}. The sparse group lasso analysis with logistic loss was conducted using phylogenomic datasets of varying sizes, ranging from 1,000 to 63,430 genes, sourced from a larger phylogenomic dataset containing DNA sequences of 63,430 (64K) genes from 363 bird species \citep{stiller2024complexity}. All sites, represented using one-hot encoding in each gene, were treated as an independent group in the analysis. The pre-processing of sequence alignments was carried out using the \textit{preprocessing} functionality of MyESL, and the largest feature file, corresponding to the 64K dataset—contained over 75 million one-hot encoded columns (genomic features).

The \textbf{sparsegl} package could successfully build ESL models for datasets with up to 5,000 genes. However, analyzing more genes exceeded the system's capacity (64 GB RAM) of a standard desktop computer. In contrast, MyESL significantly reduced computational time for building ESL models on smaller subsets (Fig 1b), requiring 10 times less computational time. Notably, MyESL was also able to build ESL models for larger data subsets, and we projected the time required for \textbf{sparsegl} for such larger datasets. Figure 1c shows that MyESL offers substantial computational time savings compared to \textbf{sparsegl}, even on regular desktops, while also requiring significantly less computational memory (Fig 1c).

\subsection{Applications of MyESL}\label{sub10}
The MyESL software provides an integrated workflow, including preprocessing sequence data, building ESL models for datasets, and post-processing ESL model outputs. The outputs generated by ESL models have a wide range of comparative and functional genomics applications. We highlighted one recent application of MyESL in identifying fragile clades in an inferred phylogeny from phylogenomic data.  

\subsubsection*{Discovering fragile clades in inferred phylogeny}
We analyzed a fungus dataset containing amino acid sequence alignments of $1,233$ genes from $86$ yeast species \citep{Shen2016-ob, Shen2017-ax}. We built an ESL model for a clade ($44$ species), where all species received a $+1$ label, while the remaining $42$ species were assigned $-1$. The combined sequence alignments from all genes contained $609,013$ sites, and the total number of bit columns was $4,105,444$, distributed among $1,233$ groups. We used weighted class balance and set the position and group sparsity parameter values at $0.1$ and $0.2$, respectively. MyESL took $5.25$ minutes to read and pre-process input datasets and less than $1$ minute to build the ESL model and process result files. The peak memory usage for this analysis was $1.3$ GB. 

The resulting ESL model contained  $33$ genes ($<3\%$), 3,745 positions ($<1\%$), and $5,608$ bit-columns ($1.3\%$). Regression coefficients were normally distributed (Fig. 1d). Two genes were much more important than others (Fig. 1e). Interestingly,  one species (\textit{A. rubescence}) in the focal clade received a low classification probability ($0.05$). Based on such an observation, \cite{sharma2024discovering} introduced a novel approach (DrPhylo) to detect fragile clades and causal gene-species combinations, implemented using MyESL. DrPhylo analysis can be conducted in MyESL using the \verb|--DrPhylo| directive. MyESL produces a two-dimensional visualization (species x genes) for this analysis that reveals causal gene-species combinations \citep{sharma2024discovering}. 

\subsection{Distributions}\label{sub11}
The source codes (Python and C++) for all custom functions used in MyESL are freely available and distributed using a GitHub repository \url{https://github.com/kumarlabgit/MyESL}. The repository contains all instructions for installing MyESL for a Python environment in Linux and performs MyESL analysis for a clade in an example phylogeny using empirical sequence alignments. We have also packaged all of these utilities of MyESL in a standalone Windows executable (.exe) file, MyESL.exe, which is also distributed through the same GitHub repository. Using this executable, we linked MyESL with DrPhylo mode to MEGA 12 \citep{kumar2024mega12} via its AppLinker interface, which made MyESL capabilities directly accessible to users with one click when the inferred phylogeny is viewed in MEGA’s Tree Explorer. 

%% file: conclusion.tex
\section{Conclusion}\label{sec:conclusion}
MyESL is an open-source, extensible, portable, efficient, and lightweight software that provides all the necessary utilities for researchers interested in using the ESL approach in molecular evolutionary and functional genomics. While a few generic packages are available for conducting sparse group lasso \citep{Yang2015-oj, Zeng2017-dl, Simon2013-nq, Klosa2020-gm, Civieta2021-wf}, they are neither optimal nor efficient for handling large phylogenomic datasets, building phylogenetic hypotheses, achieving phylogeny-aware class balancing, and domain-specific post-processing of model outputs. Some of these packages cannot be compiled into standalone applications or are proprietary, making their integration into GUI applications infeasible. MyESL overcomes these limitations.

In the above, we focussed on MyESL's use to build models for organismal relationships in a phylogeny. For example, we have used the ESL models built in MyESL to identify fragile clades and associated sequences in phylogenomics \citep{sharma2024discovering}. The use of MyESL produced highly influential positions and groups and predictive models for downstream analyses in these applications. In the future, we plan to extend the applicability of MyESL to functional and population genetic studies. Furthermore, we aim to integrate advanced sparse learning methods, such as overlapping group lasso and tree-structured lasso, to further enhance its utility for data-driven discoveries in molecular evolution and functional genomics. Furthermore, we will expand MyESL's capabilities by incorporating lasso and sparse group lasso regression with the least squared loss, enabling it to build models for continuous response variables.

%% file: conflict_of_interest.tex
\section*{Conflict of interest}
None decleared.

%% file: acknowledgements.tex
\section*{Acknowledgments}\label{sec:acknowledgements}
We thank  Alessandra Lamarca, John Allard, Hardik Sharma, Brandon Son, and Sikha Singh for their comments and for testing different versions of MyESL. 

%% file: funding.tex
\section*{Funding}\label{sec:funding}
This work was supported by a research grant from the National Institutes of Health to SK (R35GM139540-04).

%% file: data_availability.tex
\section{Data availability}
An example Fungi dataset, including amino acid sequence alignments and a phylogenetic hypothesis in a text file, is available in GitHub, which can be accessed from \url{https://github.com/kumarlabgit/MyESL}. This dataset was originally published by \cite{Shen2016-ob} and analyzed by \cite{Shen2017-ax} and \cite{sharma2024discovering}. 

%% file: main.bbl
\begin{thebibliography}{}

\bibitem[Chetverikov {\em et~al.}(2021)Chetverikov, Liao, and Chernozhukov]{Chetverikov2016-pm}
Chetverikov, D.  {\em et~al.} (2021).
\newblock On cross-validated lasso in high dimensions.
\newblock {\em Ann. Stat.}, {\bf 49}(3), 1300--1317.

\bibitem[Civieta {\em et~al.}(2021)Civieta, Aguilera-Morillo, and Lillo]{Civieta2021-wf}
Civieta, {\'A}.~M.  {\em et~al.} (2021).
\newblock asgl: A python package for penalized linear and quantile regression.
\newblock {\em arXiv:2111.00472\/}.

\bibitem[Klosa {\em et~al.}(2020)Klosa, Simon, Westermark, Liebscher, and Wittenburg]{Klosa2020-gm}
Klosa, J.  {\em et~al.} (2020).
\newblock Seagull: lasso, group lasso and sparse-group lasso regularization for linear regression models via proximal gradient descent.
\newblock {\em BMC bioinformatics\/}, {\bf 21}(1), 407.

\bibitem[Kumar and Sharma(2021)Kumar and Sharma]{Kumar2021-rp}
Kumar, S. and Sharma, S. (2021).
\newblock Evolutionary sparse learning for phylogenomics.
\newblock {\em Mol. Biol. Evol.}, {\bf 38}(11), 4674--4682.

\bibitem[Kumar {\em et~al.}(2024)Kumar, Stecher, Suleski, Sanderford, Sharma, and Tamura]{kumar2024mega12}
Kumar, S.  {\em et~al.} (2024).
\newblock Mega12: Molecular evolutionary genetic analysis version 12 for adaptive and green computing.
\newblock {\em Molecular Biology and Evolution\/}, page msae263.

\bibitem[Liang {\em et~al.}(2022)Liang, Cohen, Heinsfeld, Pestilli, and McDonald]{liang2022sparsegl}
Liang, X.  {\em et~al.} (2022).
\newblock sparsegl: An r package for estimating sparse group lasso.
\newblock {\em arXiv preprint arXiv:2208.02942\/}.

\bibitem[Liu and Ye(2010)Liu and Ye]{Liu2010-fo}
Liu, J. and Ye, J. (2010).
\newblock {Moreau-Yosida} regularization for grouped tree structure learning.
\newblock In {\em Proceedings of the 23rd International Conference on Neural Information Processing Systems - Volume 2\/}, pages 1459--1467. Curran Associates Inc., NY.

\bibitem[Liu {\em et~al.}(2011)Liu, Ji, and Ye]{Liu2011-ap}
Liu, J.  {\em et~al.} (2011).
\newblock {SLEP}: Sparse learning with efficient projections.
\newblock \textit{Note[Online].} 6:491.
\newblock \url{http://yelabs.net/software/SLEP/manual.pdf}.

\bibitem[Mendez-Civieta {\em et~al.}(2021)Mendez-Civieta, Aguilera-Morillo, and Lillo]{mendez2021agl}
Mendez-Civieta, A.  {\em et~al.} (2021).
\newblock Adaptive sparse group lasso in quantile regression.
\newblock {\em Advances in Data Analysis and Classification\/}, {\bf 15}(3), 547--573.

\bibitem[Qiao {\em et~al.}(2017)Qiao, Zhang, Su, and Lu]{qiao2017systematic}
Qiao, L.  {\em et~al.} (2017).
\newblock A systematic review of structured sparse learning.
\newblock {\em Front. Inform. Technol. Electron. Eng\/}, {\bf 18}(4), 445--463.

\bibitem[Sanderson and Curtin(2016)Sanderson and Curtin]{Sanderson2016-fr}
Sanderson, C. and Curtin, R. (2016).
\newblock Armadillo: a template-based c\texttt{++} library for linear algebra.
\newblock {\em J. Open Source Softw.}, {\bf 1}(2), 26.

\bibitem[Sharma and Kumar(2024)Sharma and Kumar]{sharma2024discovering}
Sharma, S. and Kumar, S. (2024).
\newblock Discovering fragile clades and causal sequences in phylogenomics by evolutionary sparse learning.
\newblock \textit{Mol. Biol. Evol. (revision in review)}.
\newblock \url{https://doi.org/10.1101/2024.04.26.591378}.

\bibitem[Shen {\em et~al.}(2016)Shen, Zhou, Kominek, Kurtzman, Hittinger, and Rokas]{Shen2016-ob}
Shen, X.~X.  {\em et~al.} (2016).
\newblock Reconstructing the backbone of the saccharomycotina yeast phylogeny using genome-scale data.
\newblock {\em G3: Genes, Genomes, Genetics\/}, {\bf 6}(12), 3927--3939.

\bibitem[Shen {\em et~al.}(2017)Shen, Hittinger, and Rokas]{Shen2017-ax}
Shen, X.-X.  {\em et~al.} (2017).
\newblock Contentious relationships in phylogenomic studies can be driven by a handful of genes.
\newblock {\em Nat. Eco. Evol.}, {\bf 1}(5), 126.

\bibitem[Simon {\em et~al.}(2013)Simon, Friedman, Hastie, and Tibshirani]{Simon2013-nq}
Simon, N.  {\em et~al.} (2013).
\newblock A sparse-group lasso.
\newblock {\em J. Comput. Graph. Stat.}, {\bf 22}(2), 231--245.

\bibitem[Simon {\em et~al.}(2018)Simon, Friedman, Hastie, Tibshirani, and Simon]{simon2018SGL}
Simon, N.  {\em et~al.} (2018).
\newblock Package ‘sgl’.
\newblock {\em R package version\/}, {\bf 1}.

\bibitem[Stiller {\em et~al.}(2024)Stiller, Feng, Chowdhury, Rivas-Gonz{\'a}lez, Duch{\^e}ne, Fang, Deng, Kozlov, Stamatakis, Claramunt, {\em et~al.}]{stiller2024complexity}
Stiller, J.  {\em et~al.} (2024).
\newblock Complexity of avian evolution revealed by family-level genomes.
\newblock {\em Nature\/}, {\bf 629}(8013), 851--860.

\bibitem[Tamura {\em et~al.}(2021)Tamura, Stecher, and Kumar]{tamura2021mega11}
Tamura, K.  {\em et~al.} (2021).
\newblock Mega11: molecular evolutionary genetics analysis version 11.
\newblock {\em Mol. Biol. Evol.}, {\bf 38}(7), 3022--3027.

\bibitem[Tibshirani(1996)Tibshirani]{Tibshirani1996-zs}
Tibshirani, R. (1996).
\newblock Regression shrinkage and selection via the lasso.
\newblock {\em J. Stat. Soc. Series B Stat. Methodol.}, {\bf 58}(1), 267--288.

\bibitem[Xu and Goodacre(2018)Xu and Goodacre]{Xu2018-sr}
Xu, Y. and Goodacre, R. (2018).
\newblock On splitting training and validation set: A comparative study of {Cross-Validation}, bootstrap and systematic sampling for estimating the generalization performance of supervised learning.
\newblock {\em J Anal Test\/}, {\bf 2}(3), 249--262.

\bibitem[Yang and Zou(2015)Yang and Zou]{Yang2015-oj}
Yang, Y. and Zou, H. (2015).
\newblock A fast unified algorithm for solving group-lasso penalize learning problems.
\newblock {\em Stat. Comput.}, {\bf 25}(6), 1129--1141.

\bibitem[Yang {\em et~al.}(2020)Yang, Zou, and Bhatnagar]{yang2020gglasso}
Yang, Y.  {\em et~al.} (2020).
\newblock gglasso: Group lasso penalized learning using a unified bmd algorithm.
\newblock {\em R package version\/}, {\bf 1}.

\bibitem[Zeng and Breheny(2016)Zeng and Breheny]{zengbiglasso}
Zeng, Y. and Breheny, P. (2016).
\newblock biglasso: Extending lasso model fitting to big data.
\newblock {\em R package version\/}, {\bf 1}.

\bibitem[Zeng and Breheny(2017)Zeng and Breheny]{Zeng2017-dl}
Zeng, Y. and Breheny, P. (2017).
\newblock The biglasso package: A memory-and computation-efficient solver for lasso model fitting with big data in r.
\newblock {\em arXiv:1701.05936\/}.

\end{thebibliography}
